\documentstyle[prl,aps,twocolumn,epsf]{revtex}
\begin{document}
\title{\large{\bf Density Modulations and Addition Spectra of
Interacting Electrons in Disordered Quantum Dots}}
\author{Paul N. Walker$^{1}$\cite{Home}, Yuval Gefen$^2$ and Gilles Montambaux$^1$}
\address{$^1$Laboratoire de Physique des Solides,
associ\'e au CNRS, Universit\'{e} Paris--Sud, 91405 Orsay, France}
\address{$^2$Department of Condensed Matter Physics, The Weizmann Institute
  of Science, 76100 Rehovot, Israel}
\date{\today}
\maketitle
 \begin{abstract}
We analyse the ground state of spinless fermions on a lattice in a
weakly disordered potential, interacting via a nearest neighbour
interaction, by applying the self-consistent Hartree-Fock
approximation. We find that charge density modulations emerge
progressively when $r_s\agt 1$, even away from half-filling, with
only short-range density correlations. Classical geometry dependent
{\it magic numbers} can show up in the addition spectrum which are remarkably
robust against quantum fluctuations and disorder averaging.
\end{abstract}

The interplay of disorder and interactions in two dimensional Fermi
systems is currently a central problem in condensed matter physics.
Mesoscopic systems provide a unique forum for analysing
ground state properties as it is possible to access the
regime $kT\ll \Delta$, where $\Delta$ is the mean single particle level
spacing. Examples include the study of low temperature persistent currents
and magnetic response in small quantum rings and dots, and low bias
measurements of the d.c. response \cite{Chang,Folk,Sivan,Simmel,Patel,Patel2}
and capacitance \cite{Ashoori} of weakly coupled quantum dots.
The latter experiments made it possible to access directly the energy
differences $\mu_N$ between ground states of $N$ and $N-1$ particles:
$\mu_N=E(N)-E(N-1)$. The addition spectrum, $\sum_i \delta(\mu-\mu_i)$,
depends sensitively on the nature of the mesoscopic ground state \cite{symdot}.

Resonant tunnelling measurements of the addition spectrum
\cite{Chang,Folk,Sivan,Simmel,Patel,Patel2}
have resulted in some interesting observations. Whilst the mean peak
spacings are well described by the constant interaction model
\cite{CImodel}, the fluctuations are not. In \cite{Sivan,Simmel} the
averaging was carried out over $N$, whereas in \cite{Patel,Patel2}
the results were also averaged over sample geometry. The experimental 
data \cite{Sivan} indicates the existence of atypical addition spectrum
spacings at certain values of $N$, suggesting that averaging over disorder
may not be equivalent to averaging over $N$ (the ergodicity principle
\cite{LesHouches} is violated). Theoretical and numerical studies
\cite{Sivan,Prus,BerkAlt,BMM,Stopa,Vallejos,BerkS,Walker,Orgad}
attempted to address various aspects of the problem.

In the capacitance experiments of Ref.\cite{Ashoori},
the measured addition spectra display {\it bunching}, an indication that
the Coulomb blockade becomes negative between one or more consecutive
electron addition peaks; in the experiment, these peaks then coalesce.
Such bunching is in direct conflict with the naive picture combining
the constant interaction model with ergodic effective single-particle
wavefunctions. It has been shown \cite{Shklovskii} that a classical
charge model can reproduce many of the observed effects, but there
is currently no quantum mechanical explanation as the experiments are
carried out at densities considered too high to form a Wigner solid
(in the case of a Coulomb bare interaction see ref.\cite{Ceperley}, for
a short-ranged potential one might expect a Wigner solid to be less stable).

Motivated by these experiments, but not attempting to reproduce specific details thereof, we analyse the nature of the ground
state by applying the self-consistent Hartree-Fock (SCHF) approximation.
We are thus able to go beyond the random phase approximation (RPA) with
perturbation theory (valid for large dimensionless conductance $g$,
and small $r_s$ \cite{BerkAlt,BMM}), whilst
considering larger systems than is feasible by exact methods.
Experimental values of $r_s>1$ have indeed been reported \cite{Sivan}.
Starting from a noninteracting model, we find that as the interaction
strength is increased such that $r_s \agt 1$ (but still too
weak to form a Wigner solid), the electron gas crosses over
to a regime where: (i) there are significant spatial density modulations;
(ii) density-density correlation functions seem to saturate, defining
only short range order; (iii)
the addition spectrum becomes strongly $N$ dependent, with
{\it magic numbers} for which $\Delta_2(N)\equiv \mu_{N+1}-\mu_N$
exhibits sharp maxima that coincide with the related classical charge model.

Recent reports of exact numerical studies
\cite{Berk96,Sivan,BerkAlt,BSivan}
have emphasised that the properties of quantum dots which are not
reproduced by effective single-particle random-matrix-like theories
are associated with the emergence of short range correlations. Our
results support this claim, but the main thrust here is related to
(iii): some deviations from RPA behaviour have a direct classical
electrostatic counterpart. The signature of the latter is not totally
washed out by quantum fluctuations even far from the Wigner
crystalisation threshold.

We consider the following tight binding Hamiltonian for spinless
fermions with periodic boundary conditions:
\begin{equation}
H = \sum_i w_i c^+_i c_i - 
t\sum_{\langle ij\rangle}c^+_{i} c_{j} +
{U\over 2} \sum_{\langle ij\rangle} c^+_i c^+_j c_j c_i
\label{H}
\end{equation}
where $\langle ij\rangle$ denotes pairs of nearest neighbours,
$w_i$ is the random on-site energy in the range $[-W/2,W/2]$, and
$t$ the hopping matrix element.
All lengths are measured in units of the lattice constant $a$,
so that $U=e^2/a$ and $t=\hbar^2/2ma^2$. For
low filling (i.e. a parabolic band) we find $r_s=U/(t\sqrt{4\pi\nu})$
\cite{rs} where $\nu=N/A$ is the filling factor, for $N$ electrons
in an area $A$. For the non-interacting system we find
$g=k_F l/2=96\pi\nu(t/W)^2$ by applying the Born approximation
(valid for $1 \ll g\ll A$); $k_F=\sqrt{4\pi\nu}$ and $l$ is the elastic
mean free path. In the capacitance measurements \cite{Ashoori}, the
$2d$ quantum dots was sandwiched between a metallic source
(heavily doped $n^+$ GaAs) and drain (Cr/Au), separated, at distances
comparable with the mean particle separation,  by tunnel barriers. 
To account for external sources of screening (taken as half planes), one
can insert a bare interaction between electrons in the dot that is
dipolar ($1/r^3$) at distances greater than the dot to gate separation
when there is only one close gate, and in the case of two close gates with
exponentially small long range interactions \cite{Ando}. Here we
model such interactions with a nearest-neighbour pair potential.

The ground state is obtained in the SCHF approximation, over a range of
densities and disorder strengths at zero magnetic field. The generalised
inverse participation ratio (GIPR) is then calculated according to the
following definition:
\begin{equation}
{\cal I}\equiv {1\over\nu^2 A}\sum_i \langle \rho({\bf r}_i)^2 \rangle
\ ,
\label{GIPR}
\end{equation}
where $\rho({\bf r}_i)$ denotes the expectation value of the total
density at the lattice site $i$.
The angle brackets correspond to an average over the disorder ensemble.
The GIPR provides a convenient measure of the degree of density
modulation: in the limit of a perfectly flat density profile it takes
the value unity, and increases for a modulated density. The maximal value
that can be obtained for the GIPR occurs when all the charge is concentrated
on only $N$ sites, in which case ${\cal I}=1/\nu$.

\begin{figure}[h]
\centerline{
\epsfysize 5cm
\epsffile{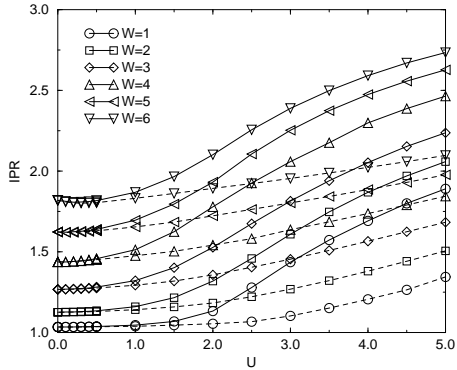}}
\caption{The GIPR is plotted for a range of disorder values as a function
of the interaction strength $U$. The system is a $16\times 16$ lattice with
n.n. (solid) and Coulomb (dashed) interactions.
We do not average over disorder, but remark that the GIPR is self averaging.
Here $\nu=1/4$, implying $r_s\approx .56 U/t$.
}
\label{fipr}
\end{figure}

The GIPR is plotted for a range of disorder strengths in Fig. 1. For
$\nu=1/4$ it
increases rapidly between $U\approx 1$ and $U\approx 5$ depending on the
disorder strength, then gives way to a weak interaction dependence for
$U\agt 5$. For comparison we also plot results for an identical system but
with bare Coulomb interactions, such that the
interaction potentials are both equal to $U$ between nearest neighbour
sites: the relative rapidity of the increase of ${\cal I}$ for
nearest neighbour interactions is clear.
The increase in the GIPR signals an increase in the spatial modulation of the
total electron density,
we shall refer to the increased density modulation at finite $U$
as a charge density modulation (CDM) \cite{CDM}.

At zero interaction we find ${\cal I}-1\sim 1/g$ for large $g$ (not shown).
Within our numerical accuracy we were unable to find a consistent size
dependence in the GIPR, suggesting that disorder is the dominant 
mechanism controlling the small to large $U$ cross-over, as seen in fig. 1
\cite{Ulimit}.

\begin{figure}[h]
\centerline{
\epsfysize 5cm
\epsffile{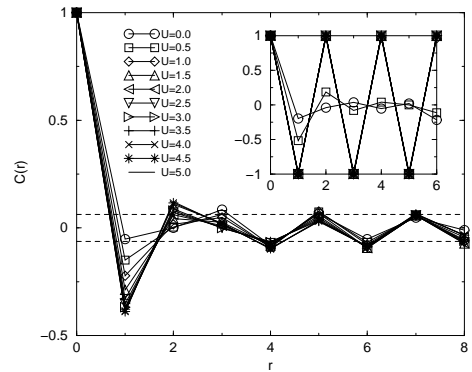}}
\caption{The density-density correlation function (see text) shows
increasing short ranged correlations as $U/t$ is increased. 
Here $W=4$, the lattice is $16*16$. Here $\nu=1/4$, implying
$r_s\approx .56 U/t$. The dotted line shows approximately one standard
deviation due to sampling, assuming that all correlations have vanished.
Inset: at half filling we find no decay of
the correlation function with distance for $U> 1$}
\label{fdcorr}
\end{figure}

The GIPR yields no information on the spatial structure of the ground
state, for which we evaluate the density-density correlation function
defined as
\begin{equation}
{\cal C}(r)=\langle \rho(r) \rho(0) \rangle_c /
\langle \rho(0)^2 \rangle_c
\ .
\label{dcorr}
\end{equation}
The subscript $c$ indicates that only connected averages are included,
and here, due to the homogeneity of the disorder averaged potential,
the correlation function depends only on the vector separation ${\bf r}$.
We only consider ${\bf r}$ to be directed along a lattice vector (1,0).
A typical result for the correlation function
is plotted in Fig. 2. As the interaction strength is increased, short
range correlation develop, and then saturate. The underlying square lattice
excludes the possibility of observing incipient Wigner crystal fluctuations,
which possess the symmetry of a triangular lattice.

Comparing figures 1 and 2, one can see that the short ranged correlations
develop over the same range of interaction strength as the rapid increase
in ${\cal I}$. We did find that on rare occasions a further
rearrangement occurs at larger interaction values, but it is not clear
whether this is a genuine effect which for larger systems would become
correspondingly less rare, or a manifestation of metastable configurations.

Let us now look at the longer range behaviour of ${\cal C}(r)$.
At half filling it has been claimed that in clean infinite lattice systems
a second order transition to a crystalline state occurs at strong interactions
\cite{Tsiper}. In disordered systems, evidence of at least short range
order has been seen in exact calculations on small systems
\cite{Berk96,BerkAlt}. Within the SCHF approximation we find no decay of
correlations.
It is well known that at half filling, nesting of the Fermi
surface leads to a $2k_F$ charge density wave instability, but
away from half filling this nesting does not occur. In fig.2 it can be
seen that, in the presence of disorder, there exists no long range order
in the SCHF ground state for $\nu=1/4$.
This however, is also true of the related classical system (i.e.
$t=0$ and $W/U\to 0$ in the Hamiltonian (\ref{H})), where one
expects the formation of a non-crystalline solid.
One way to establish whether the electrons possess solid- or liquid-
like correlations is to analyse the excited states of the system. However,
to show that classical results can provide information on the SCHF ground
state away from half filling, at least when the particle {\it packing} is
{\it compact}, we consider the appearance
of geometrical frustration, where the ground state of the classical system
contains line defects with respect to a pure crystal. These defects
lead to the disappearance of long-ranged order, but at the same time
give rise to {\it magic} filling factors where $\langle\Delta_2(N)\rangle$
exhibits large fluctuations.

This brings us to the central observation of this study, namely, the strong
geometry and filling factor dependent fluctuations that can arise in
the average addition spectrum spacing $\langle\Delta_2 (N)\rangle$
as the interaction strength is increased \cite{CLim}.
Consider first a collection of classical charges, on a square
lattice, with nearest neighbour interactions. If the lattice is a
torus with $2n\times 2m$ sites, it is possible to insert up to $2nm$
particles without incurring an energy cost. The remaining $2mn$ particles
cost an additional $4U$ to add, so that $\langle\Delta_2(N)\rangle$ displays a
peak, ${\cal O}(U)$, at $N=2nm$. If on the other hand one of the
sides (or both) of the lattice is odd (e.g. $(2n+1)\times(2m+1)$)
the maximum number of particles that can be added without nearest neighbours
is reduced. It is not difficult to see that such a maximally filled
configuration contains a line defect and long range order is lost.
In other words, the lattice of sites without nearest neighbours is
incommensurate with the underlying lattice.
In the commensurate case, the quantum system also shows a peak in
$\Delta_2(N)$ at half filling, but the nesting of the Fermi surface
in the non-interacting system
also makes this filling special. We show below that in the
incommensurate case we find that within the SCHF approximation,
remnants of the peaks at the {\it magic} filling factors in the related
classical model are visible far from the classical limit, despite the
lack of nesting.

\begin{figure}[h]
\centerline{
\epsfysize 5cm
\epsffile{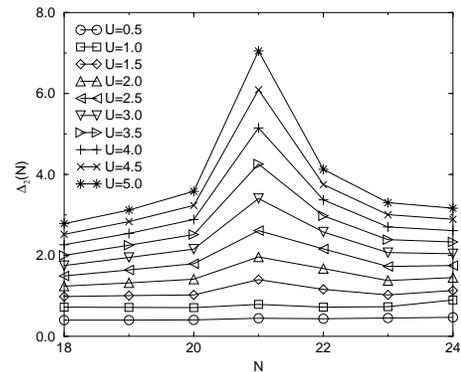}}
\caption{The addition spectrum is shown to be strongly dependent on the
number of particles in the dot, which can be understood from classical
arguments. The lattice is $7*7$, $W=2$, and the results are averaged over
400 samples.}
\label{fmn}
\end{figure}

We consider a lattice of the type $(2n+1)\times (2n+1)$ as an example,
the predictions for other incommensurate lattices are easily obtained. In
the classical limit with nearest neighbour interactions the first
$n(2n+1)$ particles can be added
with no interaction energy cost, the next $2n+1$ particles cost an additional
$2U$, and the rest cost an additional $4U$. As a result, in the classical
limit, $\langle\Delta_2(n(2n+1))\rangle$
(as well as $\langle\Delta_2((n+1)(2n+1))\rangle$) is significantly
larger than all other values of $\langle\Delta_2(N)\rangle$ \cite{CD2}.
In our calculations we include a trivial constant interaction term to
make the results easier to read. In Fig. \ref{fmn}, we
plot some typical results for $\langle\Delta_2(N)\rangle$ for a $7\times 7$ lattice,
which shows that for $U\agt 2$ remnants of this classical
effect can be seen clearly at the predicted filling $N=21$. 

Similar behaviour has been observed for other sample sizes and geometries.
Although these results correspond to a density regime where quantum
fluctuations are predominant, the structure in
$\langle\Delta_2(N)\rangle$ agrees qualitatively with that of the
classical counterpart. One might expect that extending the range of the
interaction will give rise to a more intricate classical structure, but with
correspondingly smaller amplitude which is thus more easily washed out
by quantum fluctuations. This question is left for a future study.
In previous work \cite{Shklovskii} the results of \cite{Ashoori} were
reproduced by interacting classical charges in a parabolic potential.
In this case the the source of magic numbers incorporated the existence of
topological defects in the ground state, as well as the interplay with
the confining potential. It was not {\it a priori} clear why such a
classical model should prove useful, but
our work suggests that quantum fluctuations
strong enough to destablise a Wigner solid may not completely wash
out these effects.

In summary, we show that the metallic ground state develops charge density 
modulations, controlled by the electron-electron interaction, at
densities $r_s \agt 1$ depending on disorder.
The development of the CDMs with increasing $r_s$ is more rapid for
short-range interactions, presumably because of the large gradient of
the interaction potential.
We also show that away from half filling, the CDMs are associated with
short range order only. Elsewhere \cite{Walker}, it has been demonstrated
that the existence of these CDMs result in unusual fluctuations of
$\Delta_2$ over the disorder ensemble. Finally, we demonstrate that
topological defects in the equivalent classical system occur in the CDM,
and that they result in strong filling factor and geometry dependent
fluctuations in $\Delta_2$, clearly visible for $U \agt 2$. It seems
clear that the ergodicity principle \cite{LesHouches} fails in this case,
and so disorder averaging and averaging over $N$ are not equivalent.
These results
lend support to the classical analysis of Ref.\cite{Shklovskii}, which
suggests that the behaviour seen in the experiments of Ref.\cite{Ashoori},
is due to topological defects in the classical ground state configuration.
We stress however, that bunching is not generated in the geometry
that we consider.

It is a pleasure to acknowledge discussions with A. Finkelstein, S. Levit,
B. Shklovskii and U. Sivan. This work was
supported by EU TMR fellowship ERBFMBICT961202, the
German-Israeli Foundation, the DFG as part of SFB195,
the Israel Academy of Sciences and Humanities Center for
\lq Strongly Correlated Interacting Electrons in Restricted Geometries'
and the Minerva Foundation.
Many of the calculations were made using IDRIS facilities.

%


\begin{thebibliography}{99}
\bibitem{symdot}We do not refer here to experiments on dots with discrete
symmetries and few electrons, \cite{CImodel}.
\bibitem{rs}Although $r_s$ does not have the same physical interpretation in the
presence of short ranged rather than bare Coulomb interactions, we retain
the standard definition: $\pi (r_s a_0)^2 = 1/\nu$, where $a_0$ is the
effective Bohr radius.
\bibitem{CDM}We avoid calling these density modulations
{\sl charge density waves} to eliminate
confusion with the more common usage of the latter term for long-ranged
modulations with wavevectors of $2k_F$.
\bibitem{CLim}The quantity $\langle\Delta_2 (N)\rangle$ saturates,
for very large values of $U$, to an interaction independent value.
We expect a qualitative rather than quantitative link with the classical
case because of the questionable validity of the SCHF approximation
in this limit. Here we focus on values of $U$ where the approximation
is still expected to be reliable.
\bibitem{CD2}At typical filling factors, the classical result is
$\langle\Delta_2\rangle-V\sim{\cal O}(W/A)$, where $V$ is a constant
interaction term that may be included. In the case considered, if $N$
electrons fill a {\it shell}, $\langle\Delta_2(N)\rangle-V\sim{\cal O}(U)$,
(for $U\gg W/A$).
\end{thebibliography}
\end{document}